\def\pr #1 #2 #3 {{\sl Phys. Rev.} {\bf #1}, #2 (#3)}
\def\prl #1 #2 #3 {{\sl Phys. Rev. Lett.} {\bf #1}, #2 (#3)}
\def\zp #1 #2 #3 {{\sl Z. Phys.} {\bf #1}, #2 (#3)}

\documentstyle[prl,aps,epsf]{revtex}
\begin{document}

\title{
Low Temperature Upper Critical Field Anomalies in Clean Superconductors}

\author{ G. Kotliar}
\address{
Serin Physics Laboratory \\ Rutgers University \\
Piscataway, NJ 08855-0849 \\ e-mail : kotliar@physics.rutgers.edu
}
\author{ C. M. Varma}

\address{AT\&T Bell Laboratories\\
600 Mountain Avenue, Murray Hill, NJ 07974; USA.\\
e-mail: cmv@physics.att.com}
\maketitle
\begin{abstract}
We interpret the upper-critical-field anomalies observed in some high-
temperature superconductors as resulting from the proximity to
a zero-temperature quantum critical point.
We estimate the shape of the phase boundary between the normal
and the superconducting phase by modeling the zero temperature  
critical point as the second-order endpoint of 
the first-order melting  line  of the vortex lattice.
\end{abstract}
\vskip 2 cm
\newpage

In two  recent letters,
 Mackenzie et al. \cite{mac}
and Osofsky et. al. \cite{sof}
observed that  in overdoped high  temperature superconductors
 the upper critical field can be measured at very low
temperature and displays a very steep rise as the temperature is
decreased\cite{mac}.
Mackenzie et. al. studied the single layer system
 $Tl_{2}CuO_{6+\delta}$  which is overdoped by 
incorporating excess
oxygen, while  Osofsky et al. studied $Bi_{2}SrCuO_{2}$.  

The positive curvature of the $H_{c2}$ vs. T curve and its rapid
incr anomalous low temperature cannot be explained  by 
the classical WHH (Werthamer Helfand Hohenberg) 
theory.
In fact, these anomalies  have  triggered a large number of
theoretical interpretations. Schofield and Wheatley suggested that these
anomalies are a manifestation result from the Luttinger liquid behavior
in the normal state\cite{andy}.
Alexandrov, Bratkovsky and Mott \cite{mott} account for the 
anomalous  curvature of the $H_{c2} vs T$ curves in terms of
bipolaron  superconductivity and Brandow suggested that the negative
curvature is the result of pair breaking \cite{brandow}.
Mackenzie et. al.
\cite{mac2} have suggested that the upper critical field
anomalies are related to a strong temperature
dependence of the effective mass.

In this paper we suggest a very different origin of the
upper critical field: the proximity to 
a zero temperature critical point.
We first argue, on very general grounds, that the observed 
low-temperature behavior implies the existence of a 
thermodynamic singularity at zero temperature which we
characterize in terms of two critical exponents.
Then we propose a simple model  for the relevant combination
of exponents  controlling  the shape of the  low temperature
upper critical field line which agrees
well  with the experimental observations.

The microscopic  considerations involve
the melting of the vortex lattice, a problem which
has been  a subject of
intense study \cite{vortex}.
Most of the work in this area
has concentrated  on the classical statistical mechanical
aspects of this problem
with the exception of  the recent work  
by Blatter and collaborators \cite{bla}, which carried out
a microscopic calculation of  the   quantum  and thermal
fluctuation of the vortex lattice to one loop order.

Using 
standard thermodynamic identities, one   can  relate
the slope of the upper critical field curve separating the normal
and the superconducting phase to the change in entropy and
magnetization across the transition line: 

\begin{equation}
\label{thermo}
{{ S_n  - S_s} \over { M_n - M_s}} ={{d  H_{c2} } \over { d  T}}
\end{equation}
If the superconductor-to-metal  transition  is of the
first order,
the latent heat and the magnetization jump are finite. If it is
second order,  the left  hand side of equation (\ref{thermo})
should be understood as a derivative. 
At zero temperature $S(T=0,M)=0$n the normal and the
superconducting phase. If S is
a regular function  as M approaches the upper critical
field line and the temperature tends to zero,
then  Eq.  \ref{thermo} implies that the slope of the
upper critical field curve vanishes at zero temperature.
In the experiments of Ref. \cite{mac} and \cite{sof}
the slope of the upper critical field curves diverges rather
than vanishes as the temperature approaches  zero,
implying  the existence of a singularity in the free energy at zero temperature, namely a quantum critical point.

Below the upper critical dimension, 
a second-order transition from an Abrikosov
type $II$ superconductor to a normal metal as a function of field 
is parametrized by two independent exponents  $\nu$
and $z$ which control
the divergence of a length scale and of a time scale as the critical
point is approached.
The free
energy per unit volume has a singular part
above and below the
superconducting transition which behaves as

\begin{equation}
f = A (H-H_{c2})^{\nu(d+z)}
\end{equation}

At finite temperature scaling  implies

\begin{equation}
f(T,H) = A (H-H_{c2}(T=0))^{\nu(d+z)} g (T[H-H_{c2}(T=0)]^{\nu z})
\end{equation}

The location of the upper critical field vs. temperature line is
given by the position in the temperature-field plane where the
free energy 
is singular.   If we denote the singular point  of the scaling 
function g    
by $x_c$, we obtain a connection between the
shape of the upper critical field at low temperatures and the
critical exponents:

\begin{equation}
H_{c2}(T) = H_{c2}(T=0) - x_{c} T^{\frac{1}{\nu z}} .
\end{equation}
The steep decrease in $H_{c2}$  observed experimentally is 
then a measure of the product of the 
dynamical critical exponent and the correlation length exponent of the zero-
temperature critical point.  
The scaling assumption  connects the shape of the critical line to  singularities in  other physical
quantities. We expect the linear term in the
  the normal-state specific heat to  have a singular part behaving as
$\gamma \approx (H-H_{c2})^{\nu(d-z)}$ while the a.c. susceptibility
acquires a singularity of the form 
 $ \chi \approx (H-H_{c2})^{\nu(d+z)-2}$ as we approach the upper 
critical field at low temperatures.

To estimate the exponent which determines the  shape of the upper 
critical line at low temperatures we need a  more detailed  
picture of the critical point.
Presently there is no microscopic theory of the superconductor
to metal  phase transition in the presence of a magnetic
field in  three dimensions.
To make progress we  assume that at any finite temperature, the superconductor
to normal metal transition is ( weakly) first order, an assumption
which is supported by some renormalization group calculations \cite{brezin}
and that this
first order line  ends at zero temperature in a quantum critical
point, whose existence 
is strongly suggested by  the experimental
data.
Then at any finite temperature,  we are dealing with the first
order melting transition of the vortex lattice, and the proximity
to the second order zero temperature critical point is taken into
account by using a renormalized   parameters in the determination
of the melting line.

More precisely    
we  regard 
the phase  transition
at finite temperatures as a
(weakly first order) melting of the Abrikosov lattice of
an {\it anisotropic}
three-dimensional superconductor. Then  we   estimate
the locus of
the transition line  by a modified version of the Lindemann
criterion \cite{linde}
that  takes into account the zero-temperature critical behavior near
$H_{c2}(0)$. 
Our  basic idea is that at any finite temperature we can integrate
out the quantum fluctuations to  obtain an effective action that
describes the finite temperature transition. After this integration
of the quantum fluctuations is carried out
the proximity to the quantum critical point is contained solely in
two (renormalized by quantum fluctuations ) parameters: a renormalized
stiffness and a renormalized Lindemann number.  

We assume that melting occurs when  the root  mean
square  displacement of a
vortex $\sqrt{<u^{2}>}$ due to thermal fluctuations
becomes of the order of       $d_{v}$ a quantity 
related to the distance between the normal
regions surrounding 
nearby  vortices.
The square of the distance between the centers of the vortices
$l(H)$  defined by
(${{{\sqrt 3} l^2} \over 2}={\Phi_0
\over H})$,
is  set  by the external magnetic field.
The area covered by normal region surrounding
the core is given by  
${{\sqrt(3) {l_{c}}^2} \over 2}$ with
${l_{c}}\equiv l(H_{c2}(T))$ .
The quantity    $d_{v}^2 \equiv l^2 -{l_{c}^2}$ 
vanishes as H approaches $H_{c2}$: 
\begin{equation}
\label{dv}
 d_{v}^2  \approx  l^2 
{\frac{(H_{c2}(T) -H)}{H_{c2}(T)}} .
\end{equation}
Our modification of the Lindeman criteria
is based on a 
the following picture.
After  time averaging over
many osillations of the vortex cores, 
each
vortex is associated with     an 
area  ${{{\sqrt 3} {l_{c}}^2} \over 2}+const <u^{2}> $ of normal phase.
Melting occurs when  
these normal  areas  cover the whole
sample, that is:

\begin{equation}
\label{linde}
\sqrt{<u^{2}>} = \alpha d_{v}
\end{equation}  
Here $ \alpha$ is a dimensionless  constant similar in
spirit to the Lindemann number. Based on this analogy   we expect it
to
be smaller than one. 

We stress that this is our main  assumption and is of a phenomenological
nature.
In this equation $H_{c2}(T)$ is the upper critical field computed
without taking into account the thermal fluctuations which are
responsible for the melting of the vortex lattice.

The thermal contribution to the root-mean-square 
displacement
$\sqrt{<u^{2}>}$,
is proportional to the temperature and inversely proportional to
a typical elastic constant. For the Abrikosov lattice this estimate
is complicated by the fact that some  elastic moduli are highly
non local \cite{brandt} and  softer at short wavelengths
than  at long wavelengths.
At zero wave vector the bulk modulus  $c_{11}(0) $ and the
tilt modulus  $c_{44}(0) $ 
are non critical while the elastic shear
modulus is given by $c_{66}(0) \approx (H_{c2} -H)^2$.
The expected softening of the bulk and tilt moduli
does occur at large wavevectors.For $ q \gg k_{h}$ with
$k_{h} \approx (H_{c2}-H)^{1 \over 2}$,  
$c_{44}(q) \approx (k_h/q)^2 c_{44}(0)$,
$c_{11}(q) \approx (k_h/q)^4 c_{11}(0)$ giving rise to strong 
infrared divergences in the evaluation of  $\sqrt{<u^{2}>}$.
Fortunately the relevant estimate
of the  vortex lattice mean displacement taking 
into account the nonlocality
of the bulk moduli has been carried out by 
Brandt \cite{brandt}
and by Houghton et. al. \cite{sudbo}
in the anisotropic case.
They showed that  as H approaches 
$H_{c2}$
the most divergent
contribution to the thermal displacement has the form
$
<u^{2}> = B_{1}   {l^2} \kappa^2 (\frac{H_{c2}(T)}{H_{c2}(T) -H})^{\frac{3}{2}}
{k T}
$ with 
$ B_{1} = 
(\frac{\Lambda^{2}}{c(0)_{44}c({0})_{66}})^{\frac{1}{2}}
(\frac{M_{z}}{M})^{\frac{1}{2}} \kappa \frac{1}{4\pi}$.

Here M and $M_z$ are the masses in the Landau-Ginzburg Hamiltonian,
$c(0)$ denote the elastic constants at $q=0$ , 
$\Lambda$ is
an ultraviolet cutoff of the order of the vortex lattice zone boundary
wave vector, and
$\kappa$ is the ratio of  the penetration depth to the  coherence length.
Inserting the field dependence  of the $q=0$ elastic moduli they
obtained \cite{sudbo}
\begin{equation}
\label{sudbo1}
 <u^{2}> \approx B l^2 \kappa^2
{\left(M\over M_z\right)}^{1 \over 2}  \frac{T}{({{H_{c2}-H \over
H_{c2}})}^{\frac{3}{2}}} ,
\end{equation}
where B is a constant equal to $2.26 \times 10^{-8} K^{-1}$.

To obtain the melting curve, which we denote
$H_{m}(T)$, we insert  
Eqs. (\ref{dv}) and ( \ref{sudbo1}) into the generalized
Lindemann equation (\ref{linde}).

The main difference between our equation and the usual 
Lindemann criterion which takes   $d_{v}$
to be a  non-critical constant of the order of
the magnetic length, is the critical field dependence of
 $d_{v}$ and $ <u^{2}>$.

Our assumption for (\ref{dv}) which in a more general case could
involve  an arbitrary exponent is akin to a mean-field approximation
incorporating the physics of the 
proximity to the zero-temperature critical point.
Temperature  appears explicitly in   
equations \ref{sudbo1}  and \ref{dv} and
 also implicitly in
$H_{c2}(T)$.  This temperature dependence only introduces
analytic
(proportional to $(T/T_{c})^2 $)
corrections which are  negligible
compared with the non-analytic terms that we derive below.
We therefore evaluate $H_{c2}(T)$ at zero temperature and obtain
an equation for $H_{m}(T)$ given the value of 
$H_{c2}(T=0)$.

\begin{equation}
{H_{m}(T)  \over  H_{c2}(0)} = 1  -  ({T \over T^{*}})^{\frac{2}{5}}
\end{equation}

A plot of (5) together  with the data 
of Mackenzie  et. al. \cite{mac}
in a range including 2
decades of reduced temperature is shown in Fig. 1.
The theoretical  fit used the values $H{c2} (0)= 17.4 T$ and $ T^*= 3.99 K$
We find that this estimate fits the low-temperature portion of the
data of Ref. (\cite{mac}) remarkably well. At higher temperatures and
lower fields the curves depart from  this simple power law behavior, but we
do not expect our considerations to apply far from the
zero-temperature  critical point.
At the lowest temperatures measured, there is  a hint of a crossover
to a power law with a power much closer to one. More experimental
data with smaller error bars in the low temperature region is needed
to determine wether our theory holds in the very low temperature
region or wether there is a relevant perturbation such as disorder
which is responsible for another type of critical behavior..   
The theory has a free parameter which is the value of $ \alpha $ 
in Eq. 
 (\ref{dv}). This is very similar to the parameter c in the 
Lindemann theory of melting\cite{linde}.
It determines the characteristic
temperature $T^*$ via  ${ T^* \approx {{.44 \quad  10^8  \alpha} \over
{\kappa^2 ({M_z \over M})^{1/2}}}}$.
Using  the experimental
values for the  anisotropy ratio
in $Tl_2 Ba_2 Cu O_{6+\delta}$,
$ {M \over M_z }\approx  31.6$ \cite{manako},
and  a 
Ginzburg parameter   $\kappa \approx   200$ 
\cite{wade},
we find that our value of $T^*$ corresponds to $\alpha \approx .1$.

The zero point motion gives of course a T-independent contribution
to the mean-square displacement, which can be added to the right hand
side of the Lindemann criterion.  If it is only weakly field dependent
it can  be  
considered as merely a renormalization of the core radius, 
or effectively a renormalization of $H_{c2}$ at $T=0$.
The one-loop calculation of
Blatter et al.  shows  that this is the case \cite{bla}.
We stress   that once quantum corrections are explicitly included in a renormalized value
of $H_{c2}(0)$, the elastic moduli have to be recalculated in a consistent fashion so that
they vanish when the field equals the {\it renormalized}
value of $H_{c2}$.

Other scenarios, such as the   naive application of the ordinary
Lindemann criterion, or the theory of Kosterlitz-Thouless two-
dimensional melting \cite{fisher} give an exponent of $2/3$ 
and do not agree well with the experimental data.

We now turn to the reasons why the upper-critical-field
singularity was observed only in Refs. \cite{mac} and \cite{sof}
and not in other high temperature superconductors where
fluctuations are clearly important.
First we note the the size of the upper critical field
is not anomalous if one takes into account the short
coherence length of these systems. In fact using the back-
of-the-envelope estimate for the zero temperature
upper critical field $\phi_0 = B \pi \xi^2$ 
and an
estimate of $\xi \approx 60 A^0$ we obtain $ B \approx 16 T$.  
For our considerations to be applicable  a sharp
phase transition
should take place between the
superconducting and normal phase without  an intermediate
wide crossover
region, the so called ``vortex liquid regime''.
In the  materials we discuss in this paper 
sharp  resistive transitions 
take place.
The  sharpness is due to a combination
of the purity of the samples and their strong anisotropy
which  eliminates
the  kinetic barriers responsible for the  vortex liquid phase.
Disorder and the  smaller anisotropy of YBCO cause  pinning and
vortex entanglement which result in  
a new intermediate asymptotic regime, the vortex liquid, 
broadening  the  transition between the
superconducting and the normal phase.

Our ideas relating the  $H_{c2}$ anomalies to a zero-temperature
quantum critical point  can be tested
experimentally by looking for critical behavior in  specific heat
and susceptibility measurements.

If the quantum critical point is connected to melting, we expect that
controlled addition of impurities to the sample will result in a
broadening of the transition.

The considerations in this paper are largely phenomenological
and motivated by experiments.  To justify these ideas from
microscopic considerations, one is led to the problem of the quantum melting
of the vortex lattice. If the considerations presented here are correct,
quantum effects can turn the
weakly first order finite temperature  melting transition
into a continuous transition
at  zero temperature.
\vspace{.5in}

\noindent Acknowledgment

We thank G. Blatter  B.I. Halperin,  A. Sudbo and specially
A. Schofield  for useful discussions.
Interactions with D. Vollhardt resulted in
an  improved  understanding of the material
and in a more clear  presentation.
G. K. was supported by NSF Grant No. 95-29138.2

\begin{figure}
\caption{ Upper critical field in Tesla vs temperature.
The continuous line is
given by  eq. 7 with 
$Hc2(0)= 17.4 T$ and $T^*= 3.99$
while the stars are the experimental points from Ref. [1]}
\end{figure}


\vspace{.5in}


\begin{references}

\bibitem {mac}A. P. Mackenzie et al., Phys. Rev. Lett. 71, 1238 (1993).
\bibitem {mac2}A. P. Mackenzie et al. submitted to Phys. Rev. Lett.
\bibitem{sof}  M. Osofsky et al., Phys. Rev. Lett. 71,  2315 (1993) .
\bibitem{vortex} For recent reviews see G. Blatter,  M.  Geshkenbein,
A. I. Larkin 
and V. M. Vinokur, Rev. Mod. Phys. {\bf 66}, 1125 (1994), 
and H.  Brandt preprint (1995).
\bibitem{bla} G. Blatter et. al Phys. Rev. B 40 (1994) 13013, 
G. Blatter and B. Ivlev, Phys. Rev. B 50, 10272 (1994).
\bibitem{brandow} Brandow, Int. Jour.  Mod. Phys. 8  (1994) 3859.
\bibitem{andy} A. Schofield, Phys. Rev. B 51, 11733 (1995), 
R. Dias and  J. Wheatley   Phys. Rev. B. 50, 13887 (1994).
\bibitem{mott} S. Alexandrov 
Phys. Rev. B 48, 10571 (1993).
\bibitem{brandt}  E. H. Brandt, J. Low Temp. Phys. 24, 409 (1976),
ibidem 26, 735 (1977).
\bibitem{fisher}B. Huberman and S. Doniach, Phys. Rev. Lett. 43, 950
(1979), 
D. Fisher, Phys. Rev. B 22, 1190 (1980).
\bibitem{brezin} E. Brezin, D. R. Nelson. and A. Thiaville, Phys. Rev. B 31, 7124 (1985).
\bibitem {brandt} E. H. Brandt, Phys. Rev. Lett. 63, 1106 (1989). 
\bibitem{sudbo} A. Houghton, 
 R. Pelcovits and A. Sudbo, Phys. Rev. B 40, 6763 (1989).
\bibitem{manako} T. Manako, Y. Kubo and Y. Shimakawa, Phys. Rev. B
46, 11019 (1992). 
\bibitem{linde} F. Lindemann, Z. Phys. 11, 609 (1910).
\bibitem{wade} 
A. Carrington A. P. Mackenzie, D. C. Sinclair and J. R. Cooper,
Phys. Rev. B {\bf 49} 13243 (1994), 
J. Wade et. al.,
J. of Superconductivity 7, 261 (1994).


\end{references}
\end{document}